\documentclass[aps,prd,preprintnumbers,twocolumn,groupedaddress,nofootinbib]{revtex4}
\usepackage{graphicx}
\usepackage{latexsym}
\usepackage{amsfonts}
\usepackage{amssymb}
\usepackage{amsmath}
\usepackage{slashed}
\usepackage{feynmp}
\usepackage{hyperref}
\usepackage{xspace}

\newcommand{\tev}{\ensuremath{\mathrm{\,Te\kern -0.1em V}}\xspace}
\newcommand{\gev}{\ensuremath{\mathrm{\,Ge\kern -0.1em V}}\xspace}
\newcommand{\tevt}{\ensuremath{\mathrm{Te\kern -0.1em V}}\xspace}
\newcommand{\kev}{\ensuremath{\mathrm{\,ke\kern -0.1em V}}\xspace}
\newcommand{\mev}{\ensuremath{\mathrm{\,Me\kern -0.1em V}}\xspace}

\begin{document}

\title{Particle Physics Implications for CoGeNT, DAMA, and Fermi}

\author{Matthew R.~Buckley,$^{1}$, Dan Hooper$^{1,2}$, and Tim M.P.~Tait$^{3}$}
\affiliation{$^1$Center for Particle Astrophysics, Fermi National Accelerator Laboratory, Batavia, IL 60510}
\affiliation{$^2$Department of Astronomy and Astrophysics, University of Chicago, Chicago, IL 60637}
\affiliation{$^3$Department of Physics and Astronomy, University of California, Irvine CA 92697}
\preprint{UCI-HEP-TR-2010-28,FERMILAB-PUB-10-444-PPD}
\date{\today}

\begin{abstract}

Recent results from the CoGeNT collaboration (as well as the annual modulation reported by DAMA/LIBRA) point toward dark matter with a light ($5-10 \gev$) mass and a relatively large elastic scattering cross section with nucleons ($\sigma \sim 10^{-40}$ cm$^2$). In order to possess this cross section, the dark matter must communicate with the Standard Model through mediating particles with small masses and/or large couplings. In this Letter, we explore with a model independent approach the particle physics scenarios that could potentially accommodate these signals. We also discuss how such models could produce the gamma rays from the Galactic Center observed in the data of the Fermi Gamma Ray Space Telescope. We find multiple particle physics scenarios in which each of these signals can be accounted for, and in which the dark matter can be produced thermally in the early Universe with an abundance equal to the measured cosmological density. 

\end{abstract}

\pacs{95.35.+d}

\maketitle

The CoGeNT collaboration has recently reported an excess of low energy events that are not 
consistent with known backgrounds~\cite{Aalseth:2010vx}. If interpreted as a signal of elastically 
scattering dark matter, this would imply a relatively light (5--10~$\gev$) mass for the dark matter 
(DM), and a somewhat large cross section with nucleons -- approximately (1--2)
$\times 10^{-40}~$cm$^2$~\cite{Chang:2010yk,Fitzpatrick:2010em}. Interestingly, this range of 
mass and scattering cross section is compatible with a DM explanation of the annual modulation 
reported by the DAMA/LIBRA collaboration~\cite{Drukier:1986tm}. Although null results from 
XENON10~\cite{Angle:2007uj} and CDMS-II~\cite{Kamaev:2009gp} place significant constraints 
on this interpretation, a region of parameter space remains 
open~\cite{Hooper:2010uy,Lisanti:2010qx}. Furthermore, if CoGeNT is in fact observing elastically 
scattering DM, then their rate is predicted to vary with an observable degree of annual modulation 
($\sim$10\%), providing a means with which to confirm or refute a DM interpretation of this signal in 
the coming months~\cite{Hooper:2010uy}. 

In addition, a recent analysis of the first two years of data from the Fermi Gamma Ray Space 
Telescope has revealed a flux of gamma rays concentrated around the inner $\sim$$0.5^{\circ}$ of 
the Milky Way, with a spectrum that is sharply peaked at 2-4 GeV~\cite{Hooper:2010mq}. If 
interpreted as the products of DM annihilation, this signal implies that the DM particle (which we 
will call $\chi$) has a mass between $7.3-9.2$ GeV, similar to the CoGeNT and DAMA/LIBRA value.

Following previous work~\cite{Feldman:2010ke,Fitzpatrick:2010em,Belikov:2010yi}, we
study the particle physics implications of these signals.  Working with perturbative
UV completions, we explore the set of simple theories capable of explaining the observed
signals, assuming that the CoGeNT/DAMA signals arise from the elastic scattering 
of a particle that is the majority of our Universe's DM and whose annihilations
produce the Fermi gamma ray signal.  
We begin by examining an inclusive list of mechanisms which could mediate
the DM-nucleon interaction, then turning to the Galactic Center gamma ray signal and DM
relic abundance.  Our conclusion is that the bulk of the
space of simple UV completions are likely to result in visible signals at colliders.  

The elastic scattering of a DM particle with nuclei can be written as a combination of 
spin-independent and spin-dependent couplings, and may or may not be be velocity-suppressed.
To generate the signals observed by both 
CoGeNT and DAMA/LIBRA, we must invoke spin-independent couplings, as the 
germanium used by CoGeNT contains a small quantity of isotopes with net spin, and
velocity-independent scattering because DM in the halo typically has small 
($\sim 10^{-3}$) velocity.  Significant interactions of this type may result from
scalar combinations of quarks $\bar{q} q$ or gluons $(F_{\mu \nu}^a)^2$,
a vector bilinear of quarks $\bar{q} \gamma^\mu q$, or a tensor
bilinear $\bar{q} \sigma^{\mu \nu} q$.  If we assume there are no new sources
of chiral symmetry-breaking, the scalar and tensor quark operators 
for each flavor are normalized by the corresponding quark mass.  Consequently, contributions
from the tensor operator and (from light $u/d$ quarks in the scalar operator) 
are negligible.  
We are left with 
three cases: vector interactions involving light quarks, $\sum_q m_q \bar{q} q$
(dominated by heavy quarks),
and direct contributions to the gluon operator.


First, we consider contributions to the gluon operator or colored and charged tree-level
mediators to (scalar or vector) DM-quark interactions.  Both result from integrating out a colored
heavy particle, $q^\prime$ to mediate the interaction.  In the case of a mediator for the
quark operators, the exchanged particle must be a color triplet carrying fractional electric charge.
Familiar examples of such particles are the squarks of a supersymmetric theory.
In order
to derive conservative prospects for discovery of such colored mediators, we assume
they have ${\cal O}(1)$ couplings to DM and quarks.  This provides the maximum estimate for
the mediator mass consistent with perturbation theory, and a maximum suppression of production 
at a hadron collider.

A gluon operator capable of explaining CoGeNT is already ruled out for ${\cal O}(10~\mbox{GeV})$ fermionic DM 
by a combination of
direct detection and collider searches for jets + missing energy \cite{Goodman:2010yf}.
For scalar DM, the gluon operator is consistent with collider data, 
and requires a colored mediator of 
mass $\lesssim 500$~GeV to run in the loop in order to
produce a large enough cross section.

For a $q^\prime$ mediating a vector interaction between $\chi$ and quarks, the
direct detection cross section is given by
\begin{equation}
\sigma_{p,n} = \frac{m_{p,n}^2}{\pi} \left[\left(\begin{array}{c} 2 \\ 1 \end{array}\right) 
\frac{g_{u'}}{M_{u'}^2}+ \left(\begin{array}{c} 1 \\ 2 \end{array}\right) \frac{g_{d'}}{M_{d'}^2} \right]^2, 
\label{eq:qprimecoupling}
\end{equation}
where the upper (lower) numbers refer to the cross section with protons (neutrons)
and $g_{u'}$ and $g_{d'}$ represent the $\chi$-$q$-$q'$ couplings. The mass of a given 
$q^\prime$ is heaviest when the other is decoupled from the theory. In this 
regime ({\it i.e.}~when only $u'$ or only $d'$ mediates direct detection), the signals of CoGeNT 
and DAMA/LIBRA require that $g_{q'} \approx  (M_{q^\prime}/1.2\tev)^2$. 
This places an upper limit of $1200\gev$ on the quark partner mass, assuming a 
perturbative coupling, $g_{q^\prime} \lesssim 1$.  The $q^\prime$ mediating
the scalar operator must have even smaller mass, assuming a spin flip requiring the insertion of a fermion mass.

The Tevatron currently places an lower limit of $300\gev$ on squark masses \cite{PDG}. The reach 
for fermionic quark partners profits from a four-fold increase in production cross section from the 
additional spin states.  Excluding the full mass 
range for the CoGeNT/DAMA mediator will have to wait for LHC data. With $\sqrt{s}=7\tev$, we 
estimate that $1200\gev$ squarks can be ruled out with $5-10$~fb$^{-1}$ of 
luminosity~\cite{CMSnote}. It should be noted that these estimates are
based on pair production cross section at the LHC through the
dominant $gg$ fusion channel.  We expect an additional contribution of order $5-50\%$
(for $m_{q^\prime}$ from 250-500 GeV), 
due to $q \bar{q} \rightarrow \bar{q}^\prime q^\prime$ through $t$-channel 
DM exchange, which will improve discovery prospects.



Next, we consider the case of elastic scattering mediated by a $Z'$ vector boson with a coupling of 
$g_{\chi\chi Z'}$ to DM and $g_{ffZ'}$ to a Standard Model (SM) fermion $f$. 
In terms of these couplings, the cross section 
between DM and nucleons is dominated by couplings to up and down quarks and is given by
\begin{equation}
\sigma_{p,n} \approx \frac{m_{p,n}^2 \, g_{\chi\chi Z'}^2}{\pi M_{Z'}^4} 
\left[ \left(\begin{array}{c} 2 \\ 1 \end{array}\right) g_{uuZ'}
+ \left(\begin{array}{c} 1 \\ 2 \end{array}\right) g_{ddZ'} \right]^2 .
\label{eq:Zprimecoupling}
\end{equation}
CoGeNT/DAMA needs
$g_{\chi \chi Z'} g_{q q Z'}/M^2_{Z'} \approx 0.42$ TeV$^{-2}$. 

\begin{figure}[h]

\hspace*{-0.65cm}
\includegraphics[width=1.2\columnwidth]{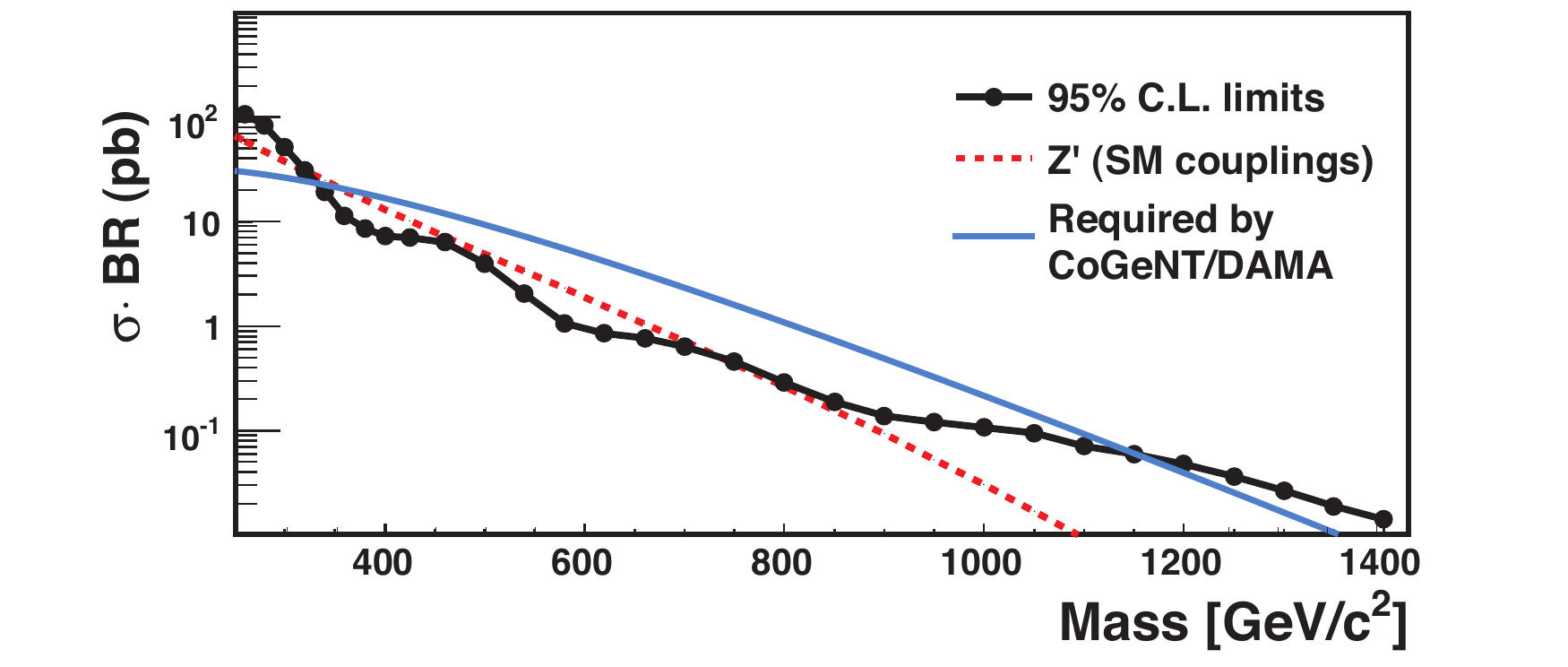}

\caption{95\% C.L.~upper limits on the $Z'$ production cross sections times the branching fraction 
to dijets~\cite{Aaltonen:2008dn} (black line and data points). The dotted red line represents the 
values predicted for a $Z'$ with SM-like couplings. The blue line indicates the prediction for
a $Z'$ mediating the CoGeNT and DAMA/LIBRA signals (for $g_{\chi \chi Z'}=1$). 
\label{fig:Zprime}}
\end{figure}

To determine conservative collider constraints, we choose $g_{\chi\chi Z'} \sim 1 \gg g_{ffZ'}$. 
A heavy $Z'$ with universal couplings to SM fermions is excluded by LEP, although a lighter 
($\sim$10 GeV) and thus more weakly coupled $Z'$ need not be leptophobic. 
The CDF collaboration has placed constraints on $Z'$ couplings to light quarks by searching for 
dijet events in 1.1 fb$^{-1}$ of data~\cite{Aaltonen:2008dn}. In Fig.~\ref{fig:Zprime}, we show how 
the constraints from that analysis impact this dark matter scenario. A $Z'$ with couplings to light 
quarks equal to those of the Standard Model (SM) $Z$ boson is ruled out over a mass range of 
approximately 350 to 800 GeV. If we set the couplings to those needed to accommodate CoGeNT 
and DAMA/LIBRA, and require that $g_{\chi\chi Z'} \leq 1$, $Z'$ masses between 350 and 1150 
GeV are excluded by CDF. 
1 fb$^{-1}$ of 7 TeV LHC data is likely to discover or exclude masses up to
 $\sim$$1600\gev$
\cite{Han:2010rf}, where the quark couplings saturate perturbativity. 
A light $Z'$ with mass below 350 GeV and small couplings could explain the data while
evading current or near future searches.
The SM $Z$ is also a viable mediator.  The $Z-\chi$ coupling necessary for CoGeNT/DAMA 
($g_{Z\chi\chi}=0.020$) is within the range allowed by measurements of the invisible $Z$ 
width ($g_{Z\chi\chi}\leq 0.023$) \cite{PDG}.



Finally, we consider DM-nucleon interactions mediated by a scalar. Naively, one might expect that 
such mediators could be either singlets or doublets of $SU(2)_L$. In the former case, the singlet 
can couple directly to gauge singlet dark matter and couple to the quarks indirectly via mixing with 
the Higgs sector. In the latter case, the doublet mediator can couple directly to the quarks, but 
requires that the dark matter itself consists of a mixture of $SU(2)_L$ singlets, doublets, and/or 
triplets (such as MSSM neutralinos). This invariably introduces heavy charged states into the dark 
sector, however, which must be $\gtrsim 90\gev$ in order to avoid limits from LEP-II~
\cite{PDG}. This requires a large splitting between the charged and neutral states, leading to
primarily singlet DM, and unacceptably small 
effective couplings to quarks for mediators with masses above $\sim$10 GeV
Therefore, we are forced to consider either light scalars with direct 
couplings to both quarks and dark matter, or a gauge singlet scalar mediator -- coupling directly 
with the dark matter and to the quarks only through mixing with the Higgs fields.

The DM elastic scattering cross section from the exchange of a scalar, $S$, is given by
$\sigma_{p,n} \approx$
\begin{equation}
\frac{4 m^4_{p,n} g^2_{\chi \chi S}}{\pi m^4_S} \bigg[\sum_{q=u,d,s} f^{(p,n)}_{T_q} \frac{g_{q q S}}{m_q} + \frac{2}{27}f^{(p,n)}_{TG}\sum_{q=c,b,t} \frac{g_{q q S}}{m_q} \bigg]^2,
\end{equation}
where $f^{(p,n)}_{T_q}$ are proportional to the matrix element, $\langle \bar{q} q \rangle$, of 
quarks in a nucleons, and $f^{(p,n)}_{TG}$ accounts for the scattering with gluons through a heavy 
quark loop. Couplings of the scalar to quarks through mixing with the SM Higgs are given by 
$g_{qqS}=g_2 m_q F_s F_{SM}/8 m_W$, where the exchanged scalar is a mixture of singlet and 
SM-like Higgs: $S=F_s H_S+ F_{SM} H_{SM}$. This yields a cross section 
\begin{equation}
\sigma_{p,n} \approx 2 \times 10^{-40}  {\rm cm}^2  \bigg(\frac{g_{\chi\chi S}}{1}\bigg)^2  \bigg(\frac{6.9 \, {\rm GeV}}{m_S}\bigg)^4  \bigg(\frac{F^2_s}{0.99}\bigg) \bigg(\frac{F^2_{SM}}{0.01}\bigg).
\end{equation}
Constraints from LEP-II ($e^+ e^- \rightarrow hZ$) rule out $F^2_{SM} \geq 0.01$
for scalar masses above about $\sim$10 GeV~\cite{Schael:2006cr}, although a very light singlet 
scalar mixed only slightly with the SM Higgs (and strongly coupled to dark matter) 
could plausibly account for the cross section implied by CoGeNT and DAMA/LIBRA.

Alternatively, we could consider the possibility of a singlet scalar that mixes with the neutral scalar 
Higgs bosons in a model with two Higgs doublets (as in many supersymmetric models, for 
example). In particular, a Higgs doublet with enhanced couplings to down-type quarks can provide 
a relatively large elastic scattering cross section with nucleons without requiring a sub-10 GeV 
mass for the scalar:
\begin{eqnarray}
\sigma_{p,n} &\approx& 2 \times 10^{-40} \, {\rm cm}^2 \\ 
&\times&\bigg(\frac{g_{\chi\chi S}}{1}\bigg)^2 \bigg(\frac{\tan \beta}{30}\bigg)^2 
\bigg(\frac{45 \, {\rm GeV}}{m_S}\bigg)^4 \bigg(\frac{F^2_s}{0.85}\bigg) 
\bigg(\frac{F^2_{D}}{0.15}\bigg), \nonumber
\end{eqnarray}
where $\tan \beta$ is the ratio of the vacuum expectation values of the two Higgs doublets and 
$F^2_D$ denotes the fraction of the singlet that is down-type Higgs doublet.

Having established several scenarios in which the signals reported by CoGeNT and DAMA/LIBRA 
could potentially arise, we now turn our attention to the process of dark matter annihilation. Ideally, 
we would like to identify cases in which both the gamma ray signal from the Galactic Center, and 
the measured cosmological density of dark matter can be accounted for. The spectrum and angular 
distribution of gamma rays from the region near the Galactic Center can be well described by a 
7.3-9.2 GeV dark matter particle which annihilates primarily to $\tau^+ \tau^-$ (possibly among 
other leptonic final states) with a cross section (to $\tau^+ \tau^-$) in the approximate range of
$3.3 \times 10^{-27}$ to $1.5 \times 10^{-26}$ cm$^3$/s \cite{Hooper:2010mq}. In order for the 
process of thermal freeze out in the early Universe to yield a density of dark matter in accordance 
with the measured value of $\Omega_{\rm DM} h^2 \approx 0.11$, the dark matter must possess an 
annihilation cross section (thermally averaged at the temperature of freeze out) of 
$\sigma v \approx 3 \times 10^{-26}$ cm$^3$/s.

We begin with the case of DM interacting through 
vector boson ($Z'$) exchange. For scalar DM particles this leads to an annihilation cross 
section that is suppressed by $v^2$, and thus cannot produce the gamma ray signal from the 
Galactic Center. If the DM is a Dirac fermion, however, the $s$-wave (non-velocity suppressed) 
portion of the cross section can be sizable and is given by~\cite{beltran}:
\begin{eqnarray}
\sigma v &=& \frac{m^2_{\chi} g^2_{\chi\chi Z'}}
{2 \pi [(M^2_{Z'}-4 m^2_{\chi})^2]+\Gamma^2_{Z'} M^2_{Z'}} \nonumber \\ 
&\times& \sum_f g^2_{ffZ'} c_f (1-m^2_f/m^2_{\chi})^{1/2} (2+m^2_f/m^2_{\chi}),
\label{zprimesigmav}
\end{eqnarray}
where $c_f=3$ for quarks and 1 for leptons. Again fixing $g_{\chi\chi Z'}=1$, we find that we require 
a value of $g_{\tau \tau Z'} \approx [0.037$ -- $0.079] \times (M_{Z'}/100\,{\rm GeV})^2$ to 
accommodate an annihilation cross section to $\tau^+\tau^-$of $3.3 \times 10^{-27}$ to 
$1.5 \times 10^{-26}$ cm$^3$/s. In comparison, couplings of $g_{q q Z'} \approx 0.0045 
\times (M_{Z'}/100\,{\rm GeV})^2$ to light quarks are needed to produce the CoGeNT and DAMA/
LIBRA signals. If we generalize these couplings to all generations, we calculate a total cross 
section of $\sigma v \approx [1.6$ -- $ 6.9] \times 10^{-26}$ cm$^3$/s, consistent with that required 
to thermally produce the measured dark matter abundance.

\begin{figure}[h]

\includegraphics[width=1.0\columnwidth]{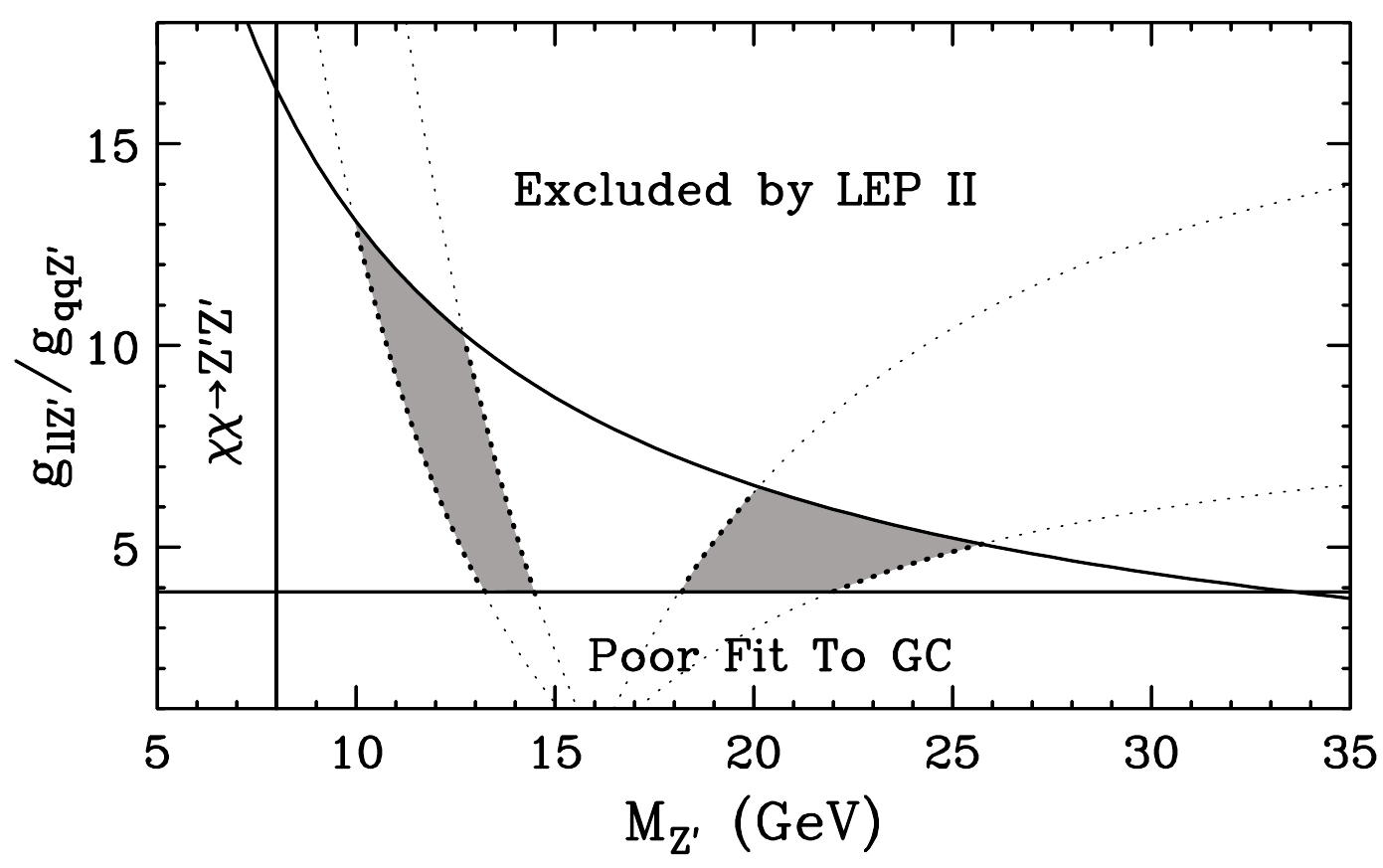}
\caption{The range of the $Z'$ masses and couplings that can accommodate the CoGeNT, DAMA/LIBRA, and Galactic Center gamma ray signals (shaded area). We have fixed the quark-$Z'$ coupling to $g_{q q Z'} \approx 0.0045 \times (M_{Z'}/100\,{\rm GeV})^2$, as required by CoGeNT and DAMA/LIBRA, set $g_{\chi \chi Z'}=1$, and the DM mass to 8 GeV. In the lower region, the fraction of annihilations to quarks is too high to provide a good fit to the Galactic Center gamma ray signal. The upper-right region is excluded by LEP-II. In the left region, the DM annihilates to $Z'$ pairs rather than fermions. The lower (upper) dashed contour denote the region in which the DM annihilation to $\tau^+ \tau^-$ is $3.3 \times 10^{-27}$ cm$^3$/s ($1.5 \times 10^{-26}$ cm$^3$/s).
\label{fig:Zprimeregions}}
\end{figure}

In Fig.~\ref{fig:Zprimeregions}, we plot the range of the $Z'$ masses and couplings that can accommodate all of the signals and constraints under consideration. 
The shaded areas are the regions in which the both  gamma ray signal from the Galactic Center and the direct detection rates reported by CoGeNT and DAMA/LIBRA can be generated.

Next on our list of potential UV completions is the exchange of a heavy colored (and 
perhaps fractionally charged) particle either at tree or loop level. 
Such objects yield annihilations only to quarks and thus will not 
be able to provide the gamma ray signal from the Galactic Center. Furthermore, the annihilation 
cross section for this process will be very small compared to that needed to provide the desired 
dark matter abundance, and thus would play only a minor role in the early Universe.

Finally, we consider dark matter annihilations mediated by a scalar. If the dark matter is a scalar, 
then this process has a large $s$-wave component, and will lead to the underproduction (over 
annihilation) of DM  in the early Universe (unless $M_S \ll 2 m_{\chi}$). Additionally, if the scalar 
gets its couplings to SM fermions through mixing with a scalar Higgs boson, then the respective 
Yukawa couplings will likely lead annihilations to $b\bar{b}$ to dominate over those to 
$\tau^+ \tau^-$, producing a gamma ray spectrum inconsistent with that observed from the Galactic 
Center (for an exception, see Ref.~\cite{Logan:2010nw}). If instead the dark matter is a fermion, 
then the annihilation cross section is $s$-wave suppressed~\cite{beltran}:
\begin{equation}
\sigma v = \frac{v^2 m^2_{\chi} g^2_{\chi\chi S}}{8 \pi [(M^2_{S}-4 m^2_{\chi})^2]+\Gamma^2_{S} M^2_{S}} \sum_f g^2_{ffS} c_f (1-m^2_f/m^2_{\chi})^{3/2}.
\end{equation}
Intriguingly, for the combination of $g_{\chi \chi S} g_{bbS}/m^2_S$ needed to generate the signal 
reported by CoGeNT and DAMA/LIBRA, this automatically leads to a thermal relic abundance near 
$\Omega_{\rm DM} h^2 \approx 0.11$~\cite{Belikov:2010yi} (assuming $M_S \gg 2 m_{\chi}$). However, due 
to the velocity suppression, no significant gamma ray signal is predicted from this 
process. Thus, to generate the observed gamma ray flux, we must introduce another annihilation 
process, for example an additional vector coupling dark matter to leptons. 
Alternatively, one could consider annihilations through the $t$-channel exchange of a particle with 
the quantum number of a tau lepton. An MSSM bino-like neutralino, for instance, annihilates to 
$\tau^+ \tau^-$ through the exchange of a stau with a cross section of 
$\sigma v \approx g_1^4 m^2_{\chi}/16\pi m^4_{\tilde{\tau}} \approx 3.7 \times 10^{-27}$ cm$^3$/s 
$\times (90\, {\rm GeV}/m_{\tilde{\tau}})^4$. One could also consider annihilations through 
a pseudoscalar, which are not $s$-wave suppressed.

In summary, we have considered an inclusive list of simple UV completions which can 
account for the signals reported by CoGeNT and DAMA/LIBRA, and for the gamma rays observed 
from the Galactic Center. We have identified a number of promising scenarios. In particular, dark 
matter consisting of a Dirac fermion which interacts with SM quarks and leptons 
through the exchange of a very weakly coupled $\sim$10-25 GeV vector boson could 
accommodate all of these signals while also producing the observed density of dark matter through 
the process of thermal freeze out in the early Universe. Alternatively, dark matter in the form of a 
fermion coupled to a singlet scalar which interacts with SM fermions through mixing with 
Higgs bosons can account for the observed elastic scattering events and yield the desired dark 
matter abundance, although an additional process would be needed to produce the observed flux 
of gamma rays.

\section*{Acknowledgements}
We would like to thank Johan Alwall, Roni Harnik, Graham Kribs, Joachim Kopp, and Adam Martin for helpful discussions.  MB and DH are supported by the US Department of Energy, including grant DE-FG02-95ER40896, and by NASA grant NAG5-10842. TT is supported by NSF grant PHY-0970171 and acknowledges the hospitality of the SLAC and Berkeley theory groups.

\end{document}